# Improved Greedy Algorithm for Set Covering Problem

Drona Pratap Chandu

*Indian Institute of Technology Roorkee, India*

**ABSTRACT :** *This paper proposes a greedy algorithm named as Big step greedy set cover algorithm to compute approximate minimum set cover. The Big step greedy algorithm, in each step selects p sets such that the union of selected p sets contains greatest number of uncovered elements and adds the selected p sets to partial set cover. The process of adding p sets is repeated until all the elements are covered. When p=1 it behaves like the classical greedy algorithm.*

## I. INTRODUCTION

An instance *(X,F)* of set covering problem (SCP) consists of a finite set *X* and a collection $F = \{S1, S2, ... Sm\}$ of subsets of *X*. The problem is to find a minimum-size subset $C \subseteq F$ such that union of members of *C* contains all the elements of *X* :

$$X = \bigcup_{s_{Kt} \in C} s_{Kt}$$

Karp[4] shown that set covering problem is NP-hard. Many approximation algorithms were proposed for set covering problem. Grossman and Wool[5] provide experimental comparison of approximation algorithms for set covering problem.

Greedy algorithms construct the solution in multiple steps by making a locally optimal decision in each step. The earlier approximation algorithms[6,7,8] use greedy heuristic. According to David[2], the advantage of greedy algorithms is that they are typically very easy to implement and hence greedy algorithms are a commonly used heuristic, even when they have no performance guarantee.

**Algorithm** GSC(*X,F*)

*X* : A finite set .

*F* : A collection of subsets of *X*.

**begin**

  $C \leftarrow \phi$

  $W \leftarrow X$

  **while**(*W* is not empty)

    Select $S \in F \setminus C$ that maximizes $|S \cap W|$

    $W \leftarrow W \setminus S$

    $C \leftarrow C \cup \{S\}$

  **end while**

  return *C*

**end**

**Figure 1. The Classical Greedy Algorithm for Set Covering Problem.**

The classical greedy algorithm for set covering problem in each step selects a set that contains greatest number of uncovered elements of *X*. The classical greedy algorithm is shown in Figure 1. The classical greedy method is explained with help of a small collection of sets in Example 1. The same collection of sets is used in Example 2 to explain Big step greedy set cover algorithm. **Example 1.** Let $X = \{a,b,c,d,e,f,g,h,i,j\}$ be the given universal set and $S = \{ \{a,b,c,d,e,f\}, \{a,b,c,g,h\}, \{d,e,f,i,j\}, \{g,h,i\}, \{j\}\}$ is the given collection of subsets of *X* . Assume labels for given sets $S1 = \{a,b,c,d,e,f\}$, $S2 = \{a,b,c,g,h\}$, $S3 = \{d,e,f,i,j\}$, $S4 = \{g,h,i\}$, $S5 = \{j\}$. Initially partial cover $C1 = \{\}$. In first step of the classical greedy algorithm, among the five sets *S1* has six uncovered elements $\{a,b,c,d,e,f\}$ and is better than the coverage of sets *S2, S3, S4,* and *S5*. So first step selects *S1* and now partial set cover *C1 = { {a,b,c,d,e,f}}*.

In second step, *S4* has three uncovered elements *{g,h,i}*, *S2* has two uncovered elements *{g,h}*, *S3* has two uncovered elements *{i,j}* and *S5* has one uncovered elements *{j}* .So second step selects *S4* and now partial cover *C1 ={{a,b,c,d,e,f},{g,h,i}}*.

In third step, *S5* has one uncovered elements *{j}*, *S2* has no uncovered elements and *S3* has one element*{j}*. So third step selects *S5* or *S3*, Let *S5* be the selected set and then *C1 = { {a,b,c,d,e,f}, {g,h,i},{j}}*.

Now *C1* contains all the elements of *X*, and *|C1|* = 3. be provided to understand easily about the paper.

## II. BIG STEP GREEDY SET COVERING ALGORITHM

The Big step greedy set cover algorithm starts with empty set cover, in each step selects *p* sets from *F* such that the union of selected *p* sets contains greatest number of uncovered elements and adds the selected *p* sets to partial set cover.

The process of adding *p* sets is repeated until all elements of *X* are covered by partial set cover. In the last step it may add less than *p* sets to avoid redundant sets .The Big step greedy set cover algorithm is shown in Figure 2.

Example 2 explains the Big step greedy set cover algorithm using the set collection used in Example 1. It can be seen that Big step greedy set cover algorithm computes smaller set cover than the classical greedy algorithm for the used collection of sets.

Results section provides performance comparison of the two algorithms using large number of instances of set covering problem.

**Example 2.** Let *X = {a,b,c,d,e,f,g,h,i,j}* is the given universal set, *S = { {a,b,c,d,e,f}, {a,b,c,g,h}, {d,e,f,i,j},{g,h,i },{j}}* is the given collection of subsets of *X* and step-size *p=2*.

Assume labels for given sets *S1 = {a,b,c,d,e,f}*, *S2 = {a,b,c,g,h}*, *S3= {d,e,f,i,j}*, *S4 ={g,h,i}*, *S5={j}*. Initially partial cover *C = {}*. As the step-size *p=2* the algorithm in each step selects two sets such that the union of two sets contain maximum number of uncovered elements.

**Algorithm** BSGSCA*(X,F,p)*

*X* : A finite set

*F* : A collection of subsets of *X*

*p* : step-size parameter of the algorithm

**begin**

  $C \leftarrow \phi$

  $W \leftarrow X$

  **while** (*W* is not empty)

    Select *S={S1,S2,....,Sp}*, $S \subseteq F\backslash C$ that maximizes | *(S1* ∪ *S2* ∪ *...* ∪ *Sp) ∩ W* |

    $V \leftarrow S1 \cup S2 \cup ... \cup Sp$

    **if** ( | *(S1* ∪ *S2* ∪ *...* ∪ *Sp) ∩ W* | = |*W*|) **then**

      $S \leftarrow$ Smallest subset *{S1,S2,... ,Sr}* of *S* such that |*(S1* ∪ *S2* ∪ *...* ∪ *Sr) ∩ W*| = |*W*|

      $V \leftarrow$ ( *S1* ∪ *S2* ∪ *...* ∪ *Sr*)

    **end if**

    $W \leftarrow W\backslash V$

    $C \leftarrow C \cup S$

  **end while**

  return *C*

**end**

**Figure 2. Big Step Greedy Set Cover Algorithm**

In the first step of algorithm, candidates are *(S1,S2) , (S1,S3) (S1,S4) (S1,S5) (S2,S3) (S2,S4) (S3,S4) (S3,S5) and (S4,S5)*, among the ten candidates *(S2,S3)* is better than all other candidates as *S2* ∪ *S3* has ten uncovered elements and is grater than that of other candidates. So the first step selects *(S2,S3)* and now partial set cover *C = { {a,b,c,g,h}, {d,e,f,i,j}}*

As all elements of *X* are covered by *C*, algorithm terminates. Finally the set over is *C = { {a,b,c,g,h} {d,e,f,i,j}}, |C| = 2*. and *C* is smaller than the set cover *C1={ {a,b,c,d,e,f}{g,h,i},{j}}* computed by the classical greedy algorithm in Example 1.

## III. EXPERIMENTAL RESULTS

In all experiments the number of elements in the universal set is 100. In the tables column labeled as |C| is the number of sets in the set collection *S* of problem instance, column labeled as "Number of Problems" is the number of problems used for performance comparison, column labeled "BSGSCA is better" is the number of problem instances for which Big step greedy set cover algorithm with *p=2* computes smaller set cover than the set cover computed by the classical greedy algorithm, and column labeled as "Greedy is better" is the number of problem instances for which the classical greedy algorithm computes smaller set cover than the set cover computed by Big step greedy set cover algorithm with *p=2*.

**Table I.** Greedy Vs Big step greedy on random set collections

| \|C\| | Problem count | BSGSCA is better | Greedy is better |
|---|---|---|---|
| 10 | 1000000 | 79996 | 30942 |
| 15 | 1000000 | 169758 | 65405 |
| 20 | 1000000 | 198858 | 72510 |
| 25 | 1000000 | 206052 | 70981 |
| 30 | 1000000 | 205607 | 64406 |
| 35 | 1000000 | 205649 | 57221 |

Table I provides details of performance comparison experiments on randomly generated problem instances. The problem instances were generated with probability of any set in the set collection containing any element from the universal set is 0.3. The value of that probability implies that the average size of subset is 3/10 of universal set.

Table II provides details of performance comparison experiments on randomly generated problem instances for the same values of problem parameters shown in Table I, but the problem instances were generated with probability of any set in the set collection containing any element from the universal set is 0.4.

**Table II.** Greedy Vs Big step greedy with *p=2* on random set collections

| |C| | Problem count | BSGSCA is better | Greedy is better |
|---|---|---|---|
| 10 | 1000000 | 141636 | 43940 |
| 15 | 1000000 | 183681 | 50238 |
| 20 | 1000000 | 176945 | 49137 |
| 25 | 1000000 | 178381 | 53346 |
| 30 | 1000000 | 178047 | 53202 |
| 35 | 1000000 | 164621 | 45344 |

Table III provides details of performance comparison on randomly generated problem instances with the probability of any set in the set collection containing any element from the universal set is 0.5.

**Table III.** Greedy Vs Big step greedy with *p=2* on random set collections

| |C| | Problem count | BSGSCA is better | Greedy is better |
|---|---|---|---|
| 10 | 1000000 | 148656 | 41899 |
| 15 | 1000000 | 177623 | 33628 |
| 20 | 1000000 | 217090 | 31212 |
| 25 | 1000000 | 222197 | 25150 |
| 30 | 1000000 | 189253 | 15337 |
| 35 | 1000000 | 140879 | 7973 |

It can be seen that Big step greedy algorithm with *p=2* computes smaller set cover than the set cover computed by the classical greedy algorithm in many cases. The difference between the performances of the algorithms is increased as the ratio of average subset size to universal set size is increased.

## IV. CONCLUSION

Experimental results show that proposed Big step greedy set cover algorithm with *p=2* computes smaller set cover than the set cover computed by the classical greedy algorithm in many cases. When step size *p* is small enough Big step greedy set cover algorithm runs in polynomial time. Big step greedy set cover algorithm is preferable than the classical greedy algorithm in scenarios where small improvement in the solution is valuable and some increment in running time is acceptable. The proposed Big step greedy method can be used for other combinatorial optimization problems.

### REFERENCES

[1] Baker, B.S., ''Approximation Algorithms for NP-Complete Problems on Planar Graphs'', Journal of the ACM, 41(1), 153–180 (1994)

[2] David P Williamson, David B. Shmoys , "The Design of Approximation Algorithms", Cambridge University Press, (Electronic web edition 2011)

[3] T. Cormen, C. Leiserson, R. Rivest: "Introduction to Algorithms", The MIT Press, (2001).

[4] Karp, R.M., "Reducibility Among Combinatorial Problems. Complexity of Computer Computations", Plenum Press (1972).

[5] Tal Grossman, Avishai Wool, "Computational experience with approximation algorithms for the set covering problem", European journal of operational research 101, 81-92, (1997).

[6] Chvatal, "A Greedy Heuristic for the set covering problem", Mathematics of Operations Research, 4/3, 223-235, (1979).

[7] Johnson, D.S., "Approximation algorithms for combinatorial problems", Journal of Computer System Science 9,256-278, (1974).

[8] Lovasz L, "On the ratio of optimal integral and fractional cover", Discrete Mathematics, 13,383-390, (1975).